\documentclass{nature}

\usepackage{xspace}
\usepackage{amsmath,amssymb}
\usepackage{hyperref}
\usepackage{graphicx}
\graphicspath{{plots/}}

\makeatletter
\let\saved@includegraphics\includegraphics
\AtBeginDocument{\let\includegraphics\saved@includegraphics}
\renewenvironment*{figure}{\@float{figure}}{\end@float}
\makeatother

\newcommand{\simba}{\mbox{{\sc Simba}}\xspace}

\title{The origin of galaxy colour bimodality in the scatter of the Stellar-to-Halo Mass Relation}

\author{Weiguang Cui$^{1}$, Romeel Dav\'e$^1$, John A. Peacock$^1$, Daniel Angl\'es-Alc\'azar$^{2,3}$, \& Xiaohu Yang$^4$}

\begin{document}

\maketitle

\begin{affiliations}
\item Institute for Astronomy, University of Edinburgh, Royal Observatory, Edinburgh EH9 3HJ, UK
\item Department of Physics, University of Connecticut, 196 Auditorium Road, U-3046, Storrs, CT 06269-3046, USA
\item Center for Computational Astrophysics, Flatiron Institute, 162 Fifth Avenue, New York, NY 10010, USA
\item Department of Astronomy, School of Physics and Astronomy, Shanghai Jiao Tong University, Shanghai, 200240, China
\end{affiliations}

\begin{abstract}
Recent observations reveal that, at a given stellar mass, blue galaxies tend to live in haloes with lower mass while red galaxies live in more massive host haloes. The physical driver behind this is still unclear because theoretical models predict that, at the same halo mass, galaxies with high stellar masses tend to live in early-formed haloes which naively leads to an opposite trend.
Here, we show that the \simba simulation quantitatively reproduces the colour bimodality in SHMR and reveals an inverse relationship between halo formation time and galaxy transition time. It suggests that the origin of this bimodality is rooted in the intrinsic variations of the cold gas content due to halo assembly bias. \simba's SHMR bimodality quantitatively relies on two aspects of its input physics: (1) Jet-mode AGN feedback, which quenches galaxies and sets the qualitative trend; and (2) X-ray AGN feedback, which fully quenches galaxies and yields better agreement with observations. The interplay between the growth of cold gas and the AGN quenching in \simba results in the observed SHMR bimodality.

\end{abstract}

\section{Introduction}
% Galaxies are formed inside dark matter haloes, whose gravitational influence attracts gas that eventually condenses and cools. This drives star formation and super-massive black hole growth, which release energy in the form of supernovae and active galactic nuclei (AGN) feedback that self-regulates galaxy stellar growth. Without such feedback, early models demonstrated that haloes would form too many stars, and massive haloes would host vigorously star-forming galaxies, in contradiction to observations\cite{Bower2006,Croton2006,Somerville2008}.

The galaxy stellar-to-halo mass relation (SHMR) represents a fundamental barometer for accretion and feedback processes in galaxy formation\cite{Wechsler2018}. At low masses, the efficiency of star formation is low, increasing towards a peak of $\sim$25\% (the ratio between stellar mass and expected baryon mass) in $M_{\rm halo}\approx 10^{12}M_\odot$ haloes, and then dropping again to higher masses\cite{Moster2013,Behroozi2013}. Its origin has been extensively explored in physical models using semi-analytic\cite{Croton2006,Somerville2008,Guo2011,Henriques2015,Croton2016} and hydrodynamic\cite{Vogelsberger2014,Schaye2015,Pillepich2018} techniques, 
which generally invoke star formation feedback to explain the SHMR below $M^\star$, and AGN feedback to explain the high-mass inefficiency\cite{Somerville2015}.

Although a bimodality in galaxy colour has long been known\cite{Strateva2001,Baldry2004,Balogh2004}, recently  it has yielded new clues into the origin of the SHMR based on observations of a bimodality in the scatter around the mean SHMR  at $z\approx 0$ (ref. \cite{More2011, Wojtak2013, Velander2014, Mandelbaum2016, PostiFall2021}): 
At a given stellar mass, for reasonably massive systems (stellar mass $M_* \gtrsim 10^{10.2} M_\odot$, or halo mass $M_{\rm halo} \gtrsim 10^{11.5} M_\odot$), red (quenched) galaxies tend to live in more massive haloes, while blue (star-forming) galaxies occupy less massive haloes. This can be reinterpreted in terms of star-forming galaxies having a higher stellar mass than quenched galaxies at a given halo mass. Note that binning in halo vs. stellar mass may result in the so-called ``inversion problem"\cite{Moster2018,Moster2020}; see \S2 in the Supplementary Information for further discussion.

Theoretical studies\cite{Zehavi2018,Matthee2017} broadly suggest that the scatter in SHMR also correlates with halo formation time, indicating the existence of {\it assembly bias}, i.e. that galaxy properties at a given halo mass depend on halo formation time.  However, linking galaxy colour to its host halo formation time is not straightforward, and contradictory results have been reported. A naive expectation is that early-formed haloes\cite{Wang2011} would host early-formed galaxies, which would quench earlier and thus be redder and less massive today\cite{Lim2016}. But observations suggest an opposite trend, that is not reproduced by theoretical models\cite{Zehavi2018,Matthee2019,Moster2020}. Therefore, the connection between the scatter in SHMR, halo assembly time, and central galaxy mass and colour is more complex than naively expected\cite{Wojtak2013,Zu2020} (see \S4 in in the Supplementary Information for further discussion), and galaxy evolution processes may play a critical role.
For instance, processes that delay star formation without invoking overly strong supernova-driven outflows could explain the observed high $M_*-M_h$ ratios of blue centrals\cite{Rodriguez-Puebla2015}. But which physical process(es) would enact this is not obvious. 

In this paper, we seek to understand the cause of the SHMR bimodality using the state of the art \simba simulation\cite{Dave2019}, described in the Methods section. 
We show that \simba quantitatively reproduces the observed SHMR bimodality for central galaxies, which is an important and fairly unique success. We examine its physical origin in terms of the {\it transition time} when a galaxy goes from fast-growing to slow-growing.  We demonstrate that the halo formation time drives a difference in the amount of cold gas within haloes, leading to halo assembly bias, while the transition time is driven by a declining cold gas content owing primarily to the onset of AGN feedback in \simba. This connects SHMR colour bimodality back to its intrinsic physical origin in the physics of AGN feedback in \simba, and gives clues as to how such feedback must by triggered in order to quench galaxies as observed.

\section{The colour bimodality in the SHMR} \label{sec:bimodality}

\begin{figure}
\includegraphics[width=0.8\textwidth]{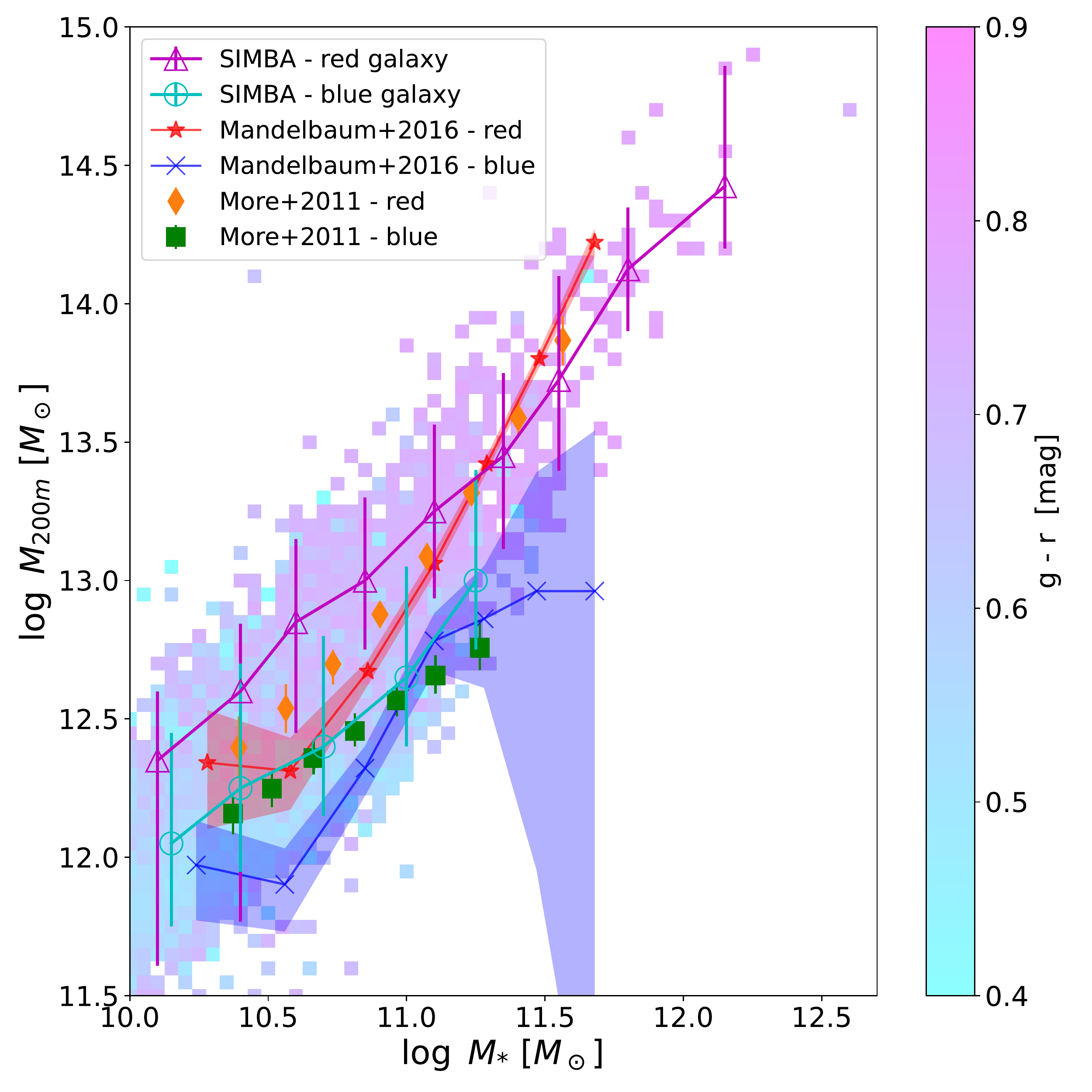}
\caption{{\bf The SHMR at $z=0$ from the \simba simulation, separated into red and blue galaxies, compared to observational results.} The simulated galaxies are binned by galaxy and halo mass, with the median $g-r$ colour in each bin shown by the colour bar. The open cyan circles and magenta triangles show median values for blue and red galaxies respectively, with error bars spanning the $16^{th} - 84^{th}$ percentiles. Using a separation based on sSFR ($\log {\rm sSFR} [{\rm Gyr}^{-1}] = -2$) yields a similar result. Observational results with halo masses based on weak lensing\cite{Mandelbaum2016} and satellite kinematics\cite{More2011} are shown with symbols and colours as indicated. Both shaded regions and error bars from the observation data indicate 1$\sigma$ error. Note that the same halo mass definition is adopted as in observations -- $M_{200m}$, the mass enclosing 200$\times$ the mean density -- and we correct for minor differences in cosmological parameters.}
\label{fig:1}
\end{figure}

Fig.~\ref{fig:1} shows that \simba well reproduces the observed bimodality in the SHMR relation, with values that are consistent with observations across the entire mass range probed, particularly when compared to the kinematic data. The galaxy colour is separated according to \cite{More2011}, see Methods section for detail. We further refer to Supplementary Fig. 1 for the galaxy colour evolution timescale in \simba. At a given stellar mass, \simba agrees that red galaxies live in larger haloes. Furthermore, \simba predicts that, at a given halo mass, blue galaxies have larger stellar masses, at least over the halo mass range $10^{11.5} \lesssim M_{\rm halo} \lesssim 10^{12.8} M_\odot$. This region is also free of the inversion problem\cite{Moster2018,Moster2020} (see Supplementary Fig. 2), since this is the interesting regime where quenching occurs. \simba's success in reproducing the observed SHMR colour bimodality motivates investigating the input physics that drives this agreement.

\begin{figure}
\includegraphics[width=\textwidth]{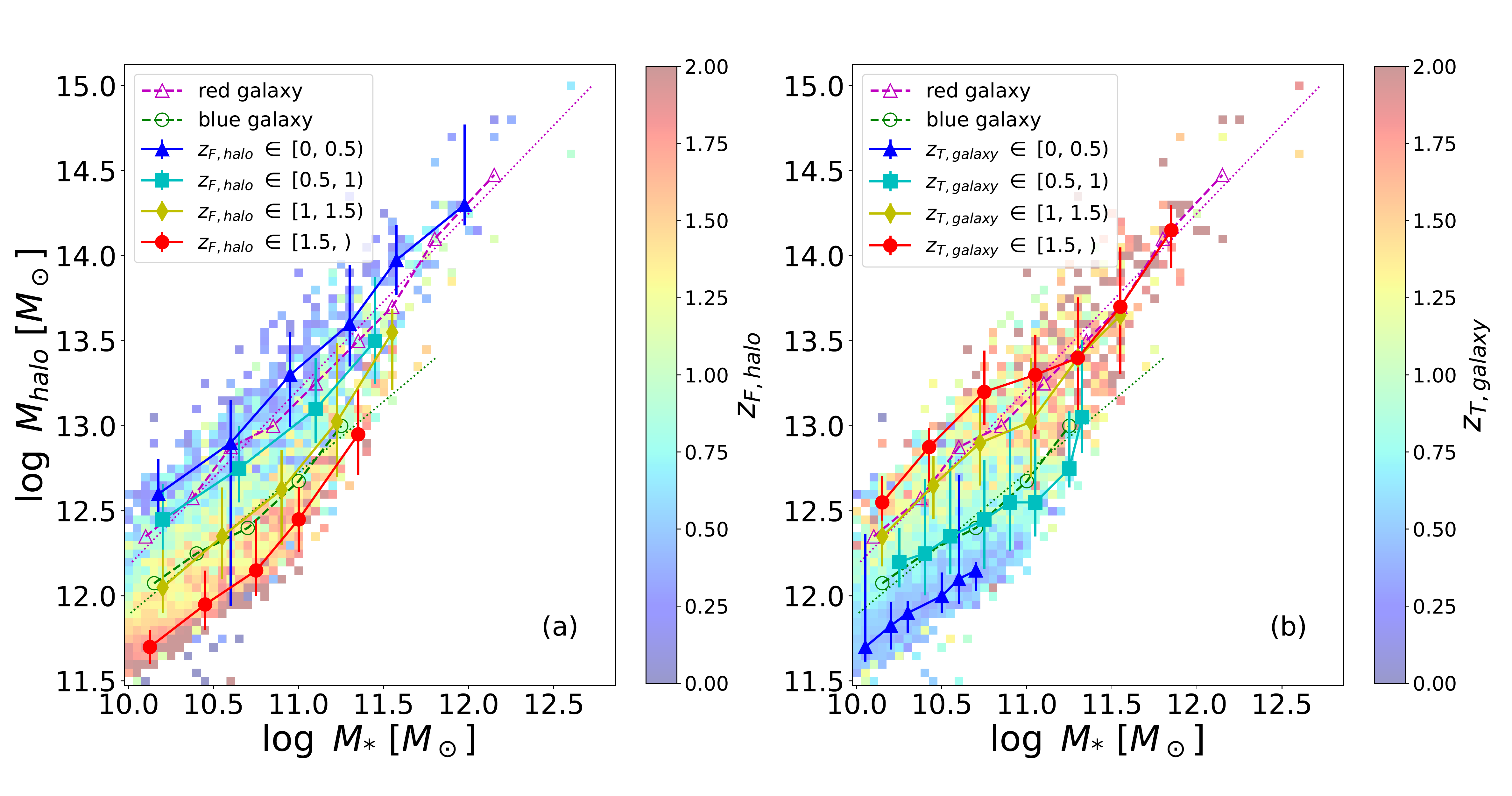} 
\caption{{\bf The SHMR from the \simba simulation, similar to Fig.~\ref{fig:1} except here colour-coded by the median halo formation redshift (a), and galaxy transition redshift (b)}. Lines of different colours show median values for the transition and formation redshift ranges indicated in the legends, with error bars showing $16^{th} - 84^{th}$ percentiles. Median lines for red and blue galaxies shown in the same symbols and colours as in Fig.~\ref{fig:1}, are also included for comparison, with dotted magenta and cyan lines showing the best linear fits to those relations: $\log M_{\rm halo} = 0.97 \log M_* -1.77$ for red galaxies, and $\log M_{\rm halo} = 1.2 \log M_* -4.23$ for blue.}
\label{fig:2}
\end{figure}

We begin by showing in Fig.~\ref{fig:2} the same SHMR as in Fig.\ref{fig:1}, but now colour-coded by halo formation time ($z_{\rm F}$) in the left panel, and galaxy transition time ($z_{\rm T}$) in the right panel.  $z_{\rm F}$ is defined as the redshift when the halo's most massive progenitor was half the present-day halo mass. $z_{\rm T}$ is defined as when the central galaxy's specific stellar growth rate exceeds 10~Gyr$^{-1}$, and characterises when the galaxy transitions from a stellar growth mode to a quiescent mode (see Methods section and Supplementary Fig. 3 for more details). 
Note that here, and in subsequent figures (except Fig.~\ref{fig:6}), a Friends-of-Friends (FoF) halo mass $M_{\rm Halo}$ instead of $M_{200m}$ is used (see Methods section for the reason). Finally,  \S2 in the Supplementary Information has a more detailed discussions on the inversion problem on $z_{\rm F}$ and ($z_{\rm T}$), but this do not influence our results in the key mass range.

Fig.~\ref{fig:2}, left panel, shows that late-formed haloes (bluer points) tend to have higher halo mass at a given $M_*$, or conversely at a given $M_{\rm halo}$ early-formed haloes host larger galaxies. \simba thus yields the same halo assembly bias trend as seen in previous results\cite{Matthee2017,Tojeiro2017,Zehavi2018,Matthee2019}. 
While it is reassuring that models broadly agree on the correlation between the scatter in galaxy stellar mass and halo formation time, this does not identify the underlying physical driver(s) of the colour bimodality. Indeed, theoretical models tend to predict galaxies with high stellar mass at a given halo mass to have a red colour\cite{Matthee2019,Moster2020}, see Section 4 in the Supplementary Information for more discussion. To understand the colour bimodality in the SHMR, we must connect the halo formation to the galaxy's mass accretion history, which sets its colour.

We parameterize the galaxy mass accretion history by the galaxy transition time $z_{\rm T}$.  In the right panel of Fig.~\ref{fig:2} we re-display the SHMR from \simba, now with the galaxies colour-coded by $z_{\rm T}$.  Clearly, $z_{\rm T}$ is also strongly correlated with the scatter in SHMR. Linking to galaxy colour bimodality, we find that blue galaxies lie in the region of the SHMR parameter space occupied by late transition galaxies ($z_T < 1$), while red galaxies tend to occupy the region where early transition galaxies ($z_T \ge 1$) are found. For more details on the connection between galaxy colour and its transition time, see Supplementary Fig. 1.

A key points is that the galaxy transition time is anti-correlated with the halo formation time: Galaxies that transition later have lower halo mass at a given $M_*$, which implies that early transition galaxies tend to live in late formed haloes, while late transition galaxies are in early formed haloes at a given halo mass. This shows that haloes that assemble quickly tend to host bluer galaxies than haloes that assemble more slowly. This inverse relationship between $z_{\rm F}$ and $z_{\rm T}$ gives rise to stellar population downsizing, in which massive haloes that assembled most recently contain the oldest stellar populations (early transition). Connecting with the observed colour bimodality, blue (late transition) galaxies tend to live in haloes with lower mass at a given stellar mass, thus in haloes with earlier formation times. Note that, at $M_{\rm halo} \gtrsim 10^{13} M_\odot$, the early transition galaxies tend to have a higher stellar mass compared the late transition ones, see Supplementary Fig. 2 for details. At lower halo mass beyond the scope of this paper, this relation may be inverted as suggested by ref. \cite{Feldmann2019}.
The inverse relation between $z_{\rm T}$ and $z_{\rm F}$ underlies how galaxy colour is connected to halo assembly bias and thus SHMR bimodality, but it still does not pinpoint the physical driver(s) in \simba.  Towards this, next we investigate the evolution of the SHMR.

\section{The evolution of the SHMR with mass growth histories} \label{sec:evolution}

\begin{figure}
\includegraphics[width=0.8\textwidth]{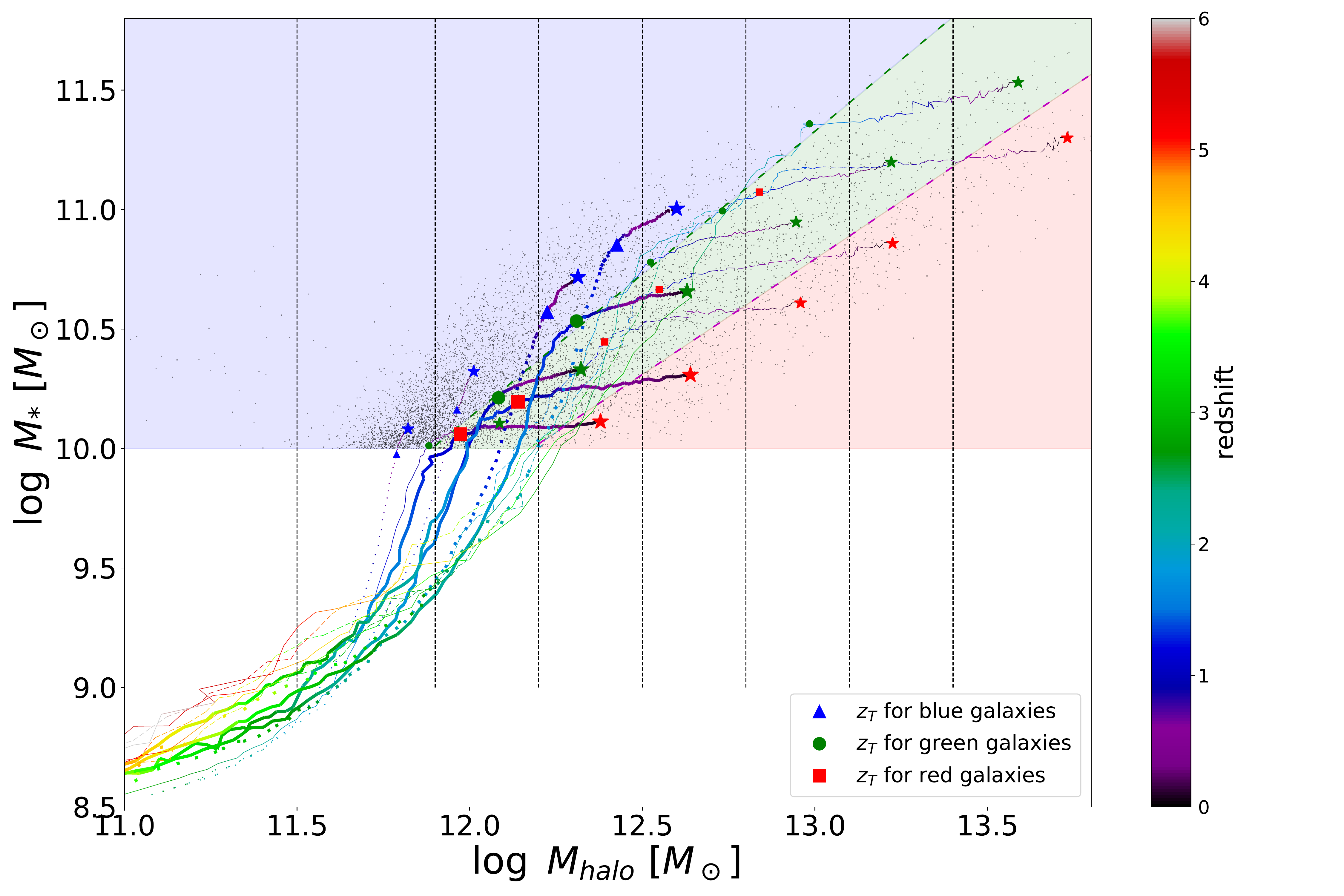} 
\caption{{\bf The evolution track of the SHMR.} The SHMR, here shown with $M_{\rm halo}$ on the $x$-axis and $M_*$ on the $y$-axis. Galaxies at $z=0$ presented in black dots, are separated into 3 different regions (blue, green and red) by the median red and blue galaxy best-fit lines from Fig.~\ref{fig:2}, and into various halo mass bins marked by the vertical dotted lines. The median stellar mass and halo mass in each zone is marked with a star symbol: blue for these lying above the fitting line for blue galaxies (early formed halo with late transition galaxies); red for these below the fitting line for red galaxies (late formed halo with early transition galaxies) and green for these in between. The median SHMR evolution track of each zone is shown by the coloured curves (dotted for blue region; solid for green region and dashed for red region), with the redshift is shown in the colour bar. The median galaxy transition times are marked with the symbols shown in legend. We consider galaxies in the 15 zones with $M_{\rm halo}>10^{11.5} M_\odot$. Note that here we use the inverted SHMR plane, putting $M_{\rm halo}$ on the $x$-axis, so that we can avoid conflating with the halo formation time dependence by keeping halo mass as the independent variable. We further use thick lines to highlight two interesting halo mass regions: $12.2 < \log M_{\rm halo} \leq 12.5$ and $12.5 < \log M_{\rm halo} \leq 12.8$ which each include all three colour regions.}
\label{fig:3}
\end{figure}

To see how the halo and galaxy growth (See Supplementary Fig. 3 for an example) manifest themselves on the SHMR plane, Fig.~\ref{fig:3} shows the median evolutionary tracks of the SHMR separated into different halo mass bins, and three colour regions (as defined from Fig.~\ref{fig:2}). Here we invert the axes since it will help illustrate our points more clearly. We note that, even in the quiescent state well after transition, galaxies can still grow via merging and residual star formation. Although we do not explicitly separate the in-situ and ex-situ growth when computing the transition time, previous work has shown that the fast stellar growth phase is dominated by in-situ growth, while ex-situ growth only becomes important for massive quiescent galaxies\cite{Hirschmann2015,Bradshaw2020}.

Prior to the transition time, the slopes of the tracks in the same halo mass bin appear anti-correlated with the final stellar masses -- the lower the final $M_*$, the higher the slope. The low slope indicates fast halo growth relative to stellar growth. If we assume that halo mass growth concurrently brings in gas, fast halo growth will bring more gas at high redshift to fuel star formation. We speculate that, owing to high early cosmic densities, this gas is expected to be cool\cite{Kere_2005,Dekel2009}. We will show that this enables the central galaxy to sustain its star formation for longer, which results in a  high stellar mass compared to galaxies at the same final halo mass that have a steeper early slope.

If we compare the curves that result in a similar final stellar mass but in different halo mass bins, it seems that they share a similar growth trend at high redshift and reach the galaxy transition time at a similar halo mass. However, the galaxies that end up in red regions have a much earlier transition time than those in blue or green regions. Therefore, after the transition time, they have a longer time to grow their halo mass while their stellar mass grows more slowly, depositing them in the lower-right portion of this SHMR plane. 
Conversely, galaxies that end up in the blue or green regions have a later transition time. This leaves less time for them to grow their halo, and they finish in the upper right (shown as the blue region) or middle (green region) of the SHMR plane.

As a side note, in Fig.~\ref{fig:3} the medians in blue region still show some post-transition growth in stellar mass, which seemingly contradicts our definition of a transition going into slow growth.  This growth is driven by a small number of galaxies that happen to have fairly high specific star formation rates such that they are close to our transition limit of 0.1.  Using a lower threshold would bring the transition time closer to $z=0$, but it will not qualitatively change our conclusions; for simplicity we keep the existing threshold for this work.

Fig.~\ref{fig:3} has interesting implications for the driver of galaxy quenching.  If we assume the transition time is correlated with the quenching time (albeit later on since galaxies are still fairly gas-rich at transition), then our results suggest that a fixed halo mass cannot be the sole driver of quenching, as has been claimed\cite{Zu2016,WangH2018}, at least for halo masses $11.5 \lesssim \log M_{\rm halo} \lesssim 12.8 M_\odot$ where \simba's SHMR does not suffer from inversion issues. This is evident because the halo masses at a given transition time for central galaxies are spread out over an order of magnitude, and do not show a single characteristic halo mass at which a galaxy enters its quiescent state. A similar conclusion on the halo mass is not a direct cause of galaxy quenching is reached by investigating the $M_{HI} - M_{halo} - M_*$ relations\cite{Guo2021}. However, halo mass still must play a key role, given that centrals in higher mass haloes have an earlier transition time and that the fraction of quenched galaxies in high mass haloes is significantly higher.  If the characteristic halo mass is still connected with galaxy quenching, one possible solution is that this halo quenching mass is redshift-dependent, i.e. a higher characteristic halo mass for quenching at higher redshifts\cite{Tinker2017,Mitra2017}. The anti-correlation between $z_{\rm T}$ and $z_{\rm F}$ and its role in setting the scatter in the SHMR indicates that the halo mass growth history plays a key role in shaping the SHMR.  In the next section, we bring in the next piece of this puzzle, the cold gas evolution, as a driver of this anti-correlation.

\section{The origin of the colour bimodality} \label{sec:origin}

As shown in previous section, at early epochs all galaxies are fast-growing and lie along a similar SHMR track. In a given (final) halo mass bin, we will demonstrate that early-forming haloes collect copious cool gas, which then begins baryon cycling via outflows to continue to provide fuel for galaxy stellar growth to later epochs -- this yields a late transition time.  Conversely, galaxies in late-forming haloes tend to have an earlier transition time, which is driven by a lower cold gas supply and exacerbated by AGN jet feedback that kicks in at low black hole accretion rates (see Methods). Furthermore, as we showed in Fig.~\ref{fig:3}, once the transition time is reached, galaxies grow primarily in halo mass with only meagre stellar growth, separating these red galaxies from the SHMR relation of the star-forming systems. To support this scenario, we first need to examine the early cold gas content difference that results from halo assembly bias.  It turns out, this will implicate \simba's AGN feedback as a key actor in SHMR bimodality.

\subsection{The intrinsic origin -- halo assembly bias and cold gas content.} \label{sec:ha}

\begin{figure}
\includegraphics[width=\textwidth]{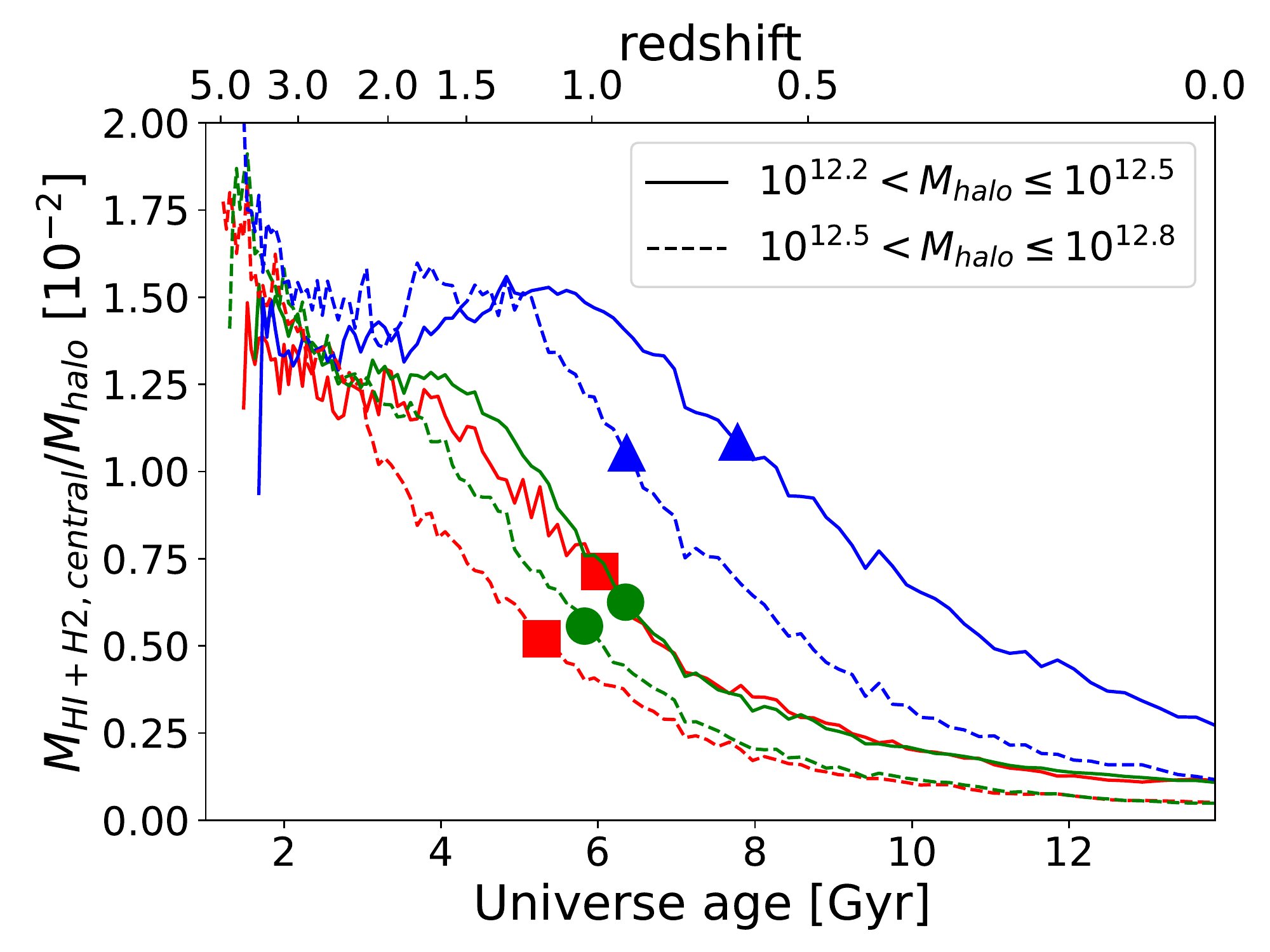} 
\caption{{\bf The evolution of the gas fraction.} The median cold gas fraction, defined as central galaxy neutral gas mass divided by halo mass, versus cosmic time. Two selected halo mass bins are shown as indicated in the legend: Solid lines are for the lower halo mass bin while the dashed lines are for the higher halo mass bin shown in Fig.~\ref{fig:3}. Different coloured lines depict results in the corresponding coloured regions in Fig.~\ref{fig:3}. The median galaxy transition times are marked as blue triangles for blue galaxies, green circles for green galaxies and red squres for red galaxies along the curves.}
\label{fig:4}
\end{figure}

Fig.~\ref{fig:4} shows the evolution of the cold gas fraction in the 6 regions in the SHMR plane as depicted in Fig.~\ref{fig:3}; here we only consider the middle two $z=0$ halo mass bins, where galaxies exist in all three coloured stellar mass regions, and are also free from the inversion problem. The cold gas fraction is defined as the HI+H$_2$ gas mass in the central galaxy divided by the host halo mass (see \S3 in the Supplementary Information).

The galaxies in the blue region are hosted by early formed haloes, i.e. more massive in halo mass at high redshifts than their present-day counterparts. We speculate that, owing to a higher early accretion rate, this higher halo mass results in a higher gas density that enhances cooling, thereby plateauing at a higher gas fraction at intermediate redshifts\cite{Tacconi_2010} when the galaxy settles into an equilibrium between inflows, outflows, and star formation\cite{Dave2012}.  This higher gas fractions persists through the transition time  (marked as a large square) as shown in Fig.~\ref{fig:4}. We suggest that this sustains the central galaxy's star formation, but also yields a high  Eddington ratio for the central black hole, which in \simba keeps the AGN feedback in a milder radiative mode that does not quench the galaxy.  The enhanced gas content at a given halo mass, driven by the early halo assembly, persists all the way to $z=0$, growing galaxies larger and keeping them bluer.

In contrast, galaxies in the red region with lower cold gas fractions are more easily able to reach the Eddington ratio limit of a few percent, triggering the jet mode AGN feedback in \simba. Jet mode has an order of magnitude higher wind speed than the radiative mode AGN feedback, and is commensurately more energetic. We suggest that this will result in significantly lower halo gas contents overall, and the gas that remains is mostly hot\cite{Dave2019,Robson2020,Appleby2021}.  The depleted hot halo gas cannot fuel star formation, leaving such late-formed haloes with redder, less massive galaxies.  Furthermore, the cold gas fractions are higher in the lower halo mass bin for all three regions, since higher halo masses host more massive galaxies with weaker winds that result in more effective conversion of gas into stars\cite{Dave2011}.

\subsection{The physical driver of galaxy transition: jet-mode AGN feedback.} \label{sec:transition}
\begin{figure}
\includegraphics[width=\textwidth]{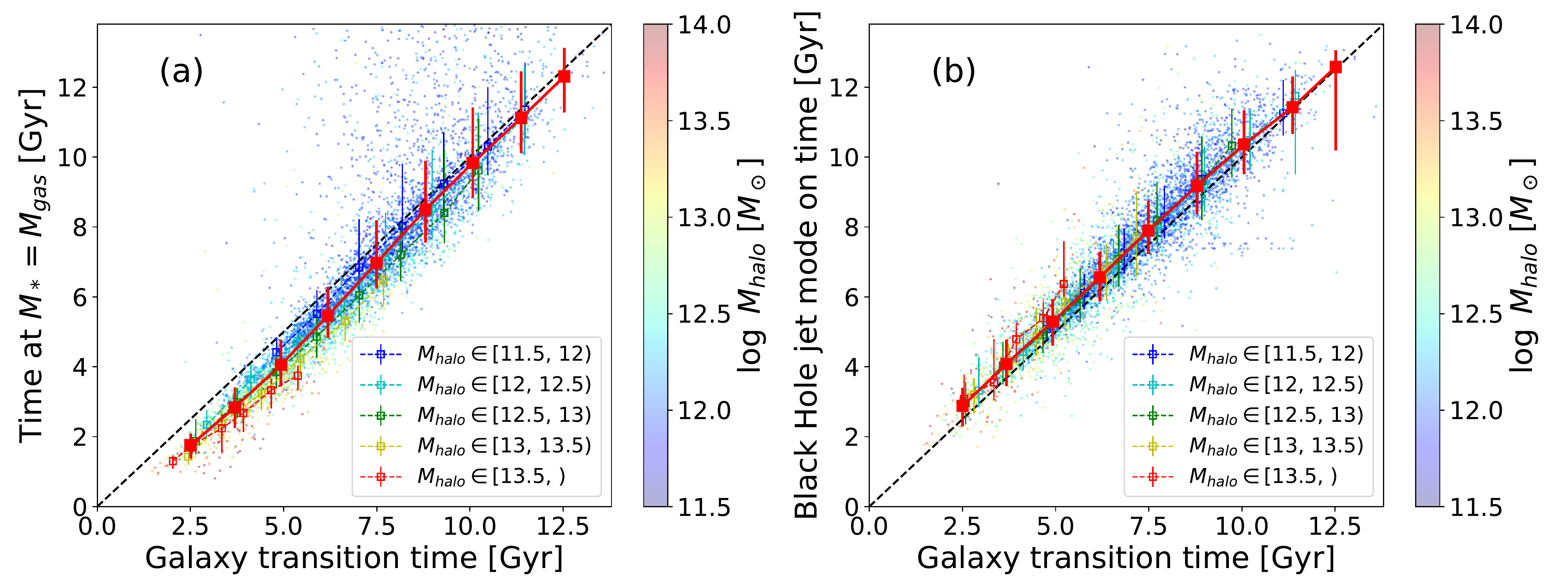} 
\caption{{\bf The correlation between galaxy transition time with crossing time (a) and AGN jet-on time(b).}  Galaxy crossing time as the y-axis in panel a, when the galaxy stellar mass is equal to its cold gas mass, versus galaxy transition time when it enters a slowly-growing phase. AGN jet-on time when jet mode AGN feedback turns on versus galaxy transition time is shown as the y-axis in panel b. Galaxies are colour coded by their halo mass at $z=0$, with the medians of different halo mass bins in coloured dashed lines. The black dashed line indicates the 1-to-1 relation. The red line with error bars shows the median values of all haloes and $16^{th}-84^{th}$ percentiles after binning in galaxy transition time.}
\label{fig:5}
\end{figure}

It is thus clear that the cold gas supply in the halo is an important driver of SHMR bimodality.  To quantify these effects and connect back to the input physics in \simba, we now introduce a new quantity, the galaxy crossing time, defined as the time when $M_* = M_{\rm gas}$.  We correlate this with the time of AGN jet turn-on, which in \simba is when the black hole's Eddington ratio drops to below 2\% (see Methods and \S3 of Supplementary Information), as well as the transition time $t_{\rm T}$ (corresponding to $z_{\rm T}$).

In Fig.~\ref{fig:5}, we show the correlations between these various times. In the left panel we show the crossing time $t_{M_* = M_{\rm gas}}$ as a function of the galaxy transition time ($t_{\rm T}$), while in the right panel we show $t_{\rm jet\, on}$ versus $t_{\rm T}$. 
Remarkably, both $t_{M_* = M_{\rm gas}}$\,-\,$t_{\rm T}$ and $t_{\rm jet\, on}$\,-\,$t_{\rm T}$ nearly follow a one-to-one relation (dashed lines in Fig.~\ref{fig:5}), confirming the driving role of these aspects in setting the galaxy transition time. As an aside, note the $\sim0.5$ Gyr delay between $t_{M_* = M_{\rm gas}}$ and the transition time for massive haloes (red and yellow dashed lines) in agreement with Supplementary Fig. 3, indicating that galaxy quenching requires a lower gas fraction than galaxy transition.  This correlation is in agreement with the expulsion of efficiently cooling gas from the CGM as a crucial step in quenching a galaxy\cite{Davies2020}.
While the galaxy SFR correlates with the cold gas mass\cite{Kennicutt2012}, the fact that $t_{\rm T}$ is proportional to $t_{M_* = M_{\rm gas}}$ is a non-trivial consequence of the upward trend of increasing cold gas mass at early times, tightly connected with halo formation time, and a downward trend of decreasing cold gas mass at late times, which can be driven by depletion by star formation, heating/outflows by feedback, and/or virial shock heating\cite{Kere_2005}. Crucially, the lower gas fraction connects to a low Eddington ratio that triggers the jet mode AGN feedback, as shown in the right panel of Fig.~\ref{fig:5}, which quenches the galaxy\cite{Dave2019}.

We have thus demonstrated that \simba's success in reproducing the SHMR scatter owes to its quenching feedback that coincides with the halo cold gas content dropping to low values, thus triggering the jet-mode AGN feedback that transitions the galaxy.  However, this turns out not to fully explain the SHMR bimodality, as we show next; one further ingredient is needed in \simba.

\subsection{Enhancing the galaxy colour separation: X-ray feedback.} \label{sec:xray}

\begin{figure}
\includegraphics[width=1\textwidth]{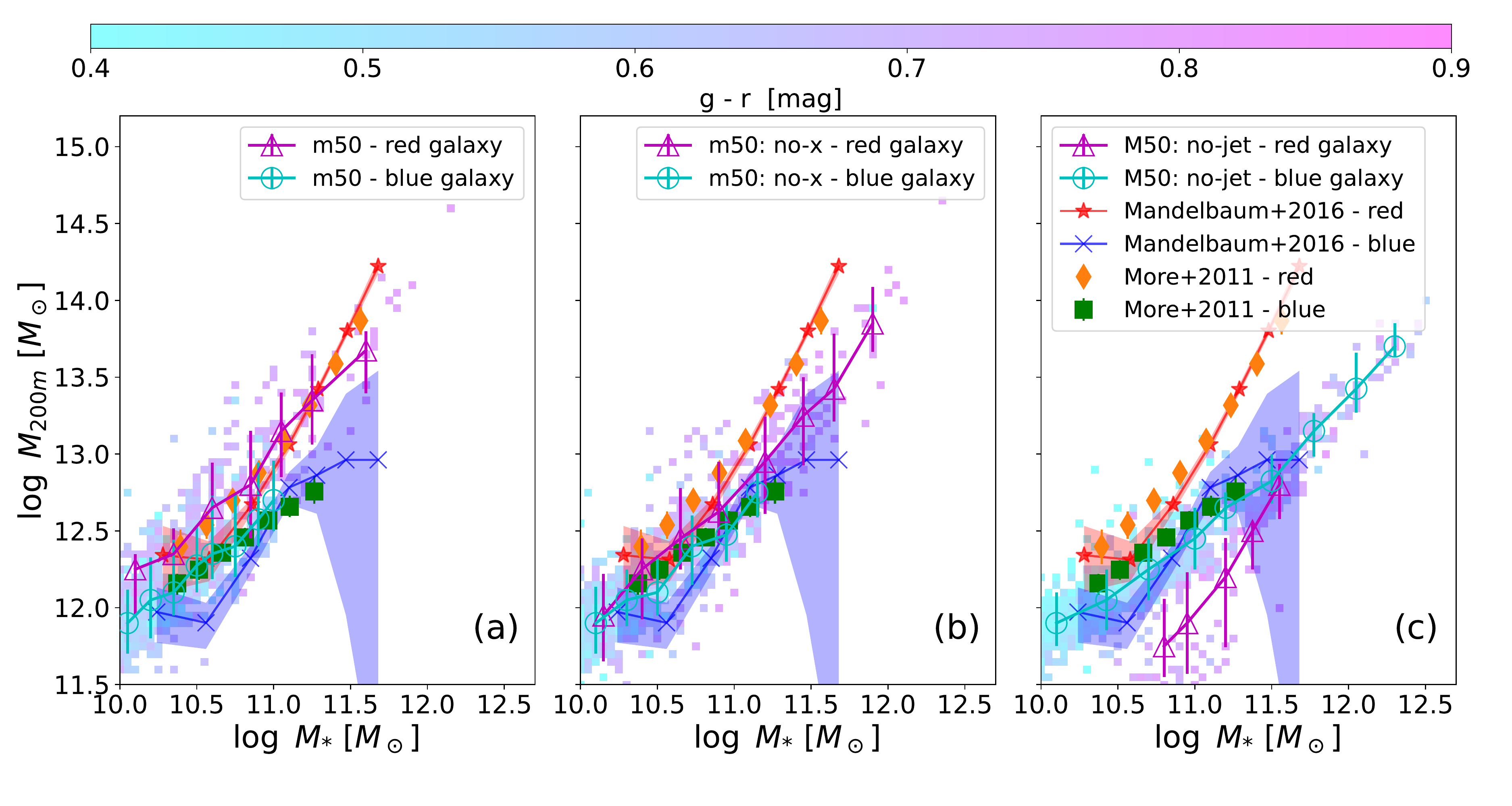}
\caption{{\bf The SHMR of three simulations with different baryon models: fiducial(a), `no-x'(b), `no-jet'(c).} SHMR subdivided into red and blue galaxies via the same colour cut as in Fig.~\ref{fig:1}, but for a smaller 50\,$h^{-1}\,{\rm Mpc}$ simulation box with the full \simba feedback (same as the default run) in panel a, an identical `no-X' run turning off only X-ray AGN feedback in panel b, and a `no-jet' run further turning off the jet-mode AGN feedback in panel c. Note that the same notations and error bars as in Fig.~\ref{fig:1} are adopted here.}
\label{fig:6}
\end{figure}

We can quantitatively assess how \simba arrives at its SHMR bimodality by examining \simba's AGN feedback variant runs.
Fig.~\ref{fig:6}, left panel, shows the SHMR as in Fig.~\ref{fig:1} from three 50\,$h^{-1}\,{\rm Mpc}$ runs with identical initial conditions: A
full \simba physics run (fiducial; left panel),  a run turning off X-ray AGN feedback (`No-X'; middle), and a run turning off both X-ray and jet mode feedback (`No-jet'; right). Note that $M_{200m}$ instead of $M_{FOF}$ halo masses are used here for a consistent comparison with the observed results. The fiducial run shows similar results to the main 100\,$h^{-1}\,{\rm Mpc}$ \simba\ run in Fig.~\ref{fig:1}, confirming that our results are not sensitive to simulation volume.

Starting with the No-jet case (right panel), we see that the median SHMR for blue galaxies (cyan line) is similar to the fiducial full-physics run, but the SHMR for red galaxies (magenta line) is drastically shifted towards higher $M_*$, so that the colour bimodality in the SHMR is now reversed!  Thus without quenching, the SHMR follows the naive expectation that galaxy growth simply follows halo growth, so late-formed haloes continue to harbor star formation and grow their stellar mass until today. This is in clear disagreement with observations.  

Turning on only jet feedback (No-X, middle panel), again the blue SHMR remains mostly unchanged, but the red galaxies' SHMR has now moved slightly above the blue.  Thus jet feedback goes the majority of the way towards establishing the SHMR colour bimodality -- but it still does not fully match observations.  This requires turning on X-ray AGN feedback (left panel), which fully quenches galaxies~\cite{Dave2019}, and moves the quenched galaxy SHMR to even somewhat lower $M_*$ and thus quantitatively reproduces the observations.

Thus the basic qualitative trend of the SHMR colour bimodality in \simba owes to jet mode AGN feedback, while quantitative agreement further requires the inclusion of X-ray AGN feedback.  This at last completes the connection between the SHMR scatter bimodality and the driving input physics in \simba.

\section{Conclusion}

We recap our scenario for the origin of the colour bimodality in the SHMR in \simba, which is successful in reproducing observations without any specific tuning. The correlation between the scatter in the SHMR and halo formation time reveals that the colour bimodality is rooted in intrinsic variation of the early cold gas content owing to halo assembly bias: at a given final halo mass, early-formed haloes have higher cold gas fractions, and vice versa. In late-formed haloes with low cold gas fractions, the galaxy transitions earlier, thus setting up an anti-correlation of the halo assembly time and galaxy transition time.  The physics driving this in \simba is primarily jet-mode AGN feedback, which kicks in at low Eddington ratios that occur when cold gas fractions are low, and quenches the galaxy. Furthermore, the X-ray AGN feedback model in \simba, which kicks in for jet-mode galaxy with a low gas fraction, removes any residual star formation and thus keeps the galaxy in quiet, leading to red colour. Once the galaxy transitions to slow-growing, the halo continues to grow, placing these red galaxies in more massive haloes by $z=0$.  Conversely, early-formed haloes accumulate significant cold gas early on, which delays the onset of quenching and sustains star formation for longer to yield bluer galaxies.

We have used the \simba simulation to illustrate how its input physics drives SHMR bimodality. But even if the details of SIMBA’s feedback models are not correct, our results are likely to be generalizable to other galaxy formation models. What \simba suggests is that in order to obtain the SHMR colour bimodality, it is necessary that early-forming haloes obtain more cold gas (which is expected from structure formation), and that they must also retain this advantage to $z=0$.  To do this, we suggest that the feedback that quenches galaxies, whatever that is (e.g. jet feedback in \simba), must operate preferentially in haloes with lower cold gas fractions.  This is critical for producing the observed trend of the SHMR bimodality.  Obtaining quantitative agreement with data requires keeping galaxies fully quenched and red; in \simba this occurs via X-ray feedback that pushes residual cold gas out.  While other models may implement quenching feedback in different ways, our results suggest that if quenching is connected to haloes that have preferentially lower cold gas to fuel star formation, the net effect is likely to be similar to that of \simba's AGN feedback modules, and will qualitatively reproduce the observed SHMR colour bimodality.  That said, our results suggest that obtaining quantitative agreement with SHMR colour bimodality provides a stringent test for models of galaxy quenching, particularly as constraints improve with upcoming facilities such as the Vera Rubin Telescope.

\section{Methods} \label{methods}
\subsection{The Simba simulation.} \label{sec:simba}
The $100\,h^{-1}\,$Mpc main \simba simulation \cite{Dave2019} is used for this study. \simba is based on the {\sc Mufasa} simulation \cite{Dave2016} with its sub-resolution star formation and stellar feedback prescriptions, with galactic wind scaling taken from the FIRE simulations\cite{Muratov2015, AA2017b}. \simba further includes two models of black hole growth prescriptions: the torque-limited accretion model from cold gas \cite{AA2015,AA2017} and Bondi accretion from hot gas. AGN feedback is modelled via kinetic bipolar outflows, the strength of which depends on the BH accretion rate, separated into three modes: a `radiative mode' at high Eddington ratio to drive multi-phase winds at velocities $\sim 10^3$~km~s$^{-1}$; a `jet mode' at low Eddington ratios $f_{\rm Edd} < 0.2$, where AGN drive hot gas in collimated jets at velocities $\sim 10^4$~km~s$^{-1}$; a X-ray heating from black holes which aims to represent the momentum input from hard photons radiated off from the accretion disk~\cite{Choi2012}, only initiated after jet mode feedback is on. This simulation is tuned to reproduce the stellar mass function up to redshift $\sim 6$. It furthermore reproduces the quenched fractions and the main sequence of star-forming galaxies at various $M_*$ (ref. \cite{Dave2019,Katsianis2020}), the observed black hole mass -- galaxy stellar mass and velocity dispersion relations\cite{Thomas2019}, the observed dust mass function at $z=0$ (ref. \cite{Li2019}), and many other galaxy properties\cite{Appleby2020}. 

In addition, a series of smaller box (50\,$h^{-1}\,{\rm Mpc}$) simulations with the same resolution as the $100\,h^{-1}\,$Mpc main \simba simulation, but different feedback models, are also used to investigate the effects of different AGN feedback in this study: a fiducial run which shares everything with the $100\,h^{-1}\,$Mpc run; a `No-X' run which only turns off X-ray feedback; a `No-jet' run which turns off both X-ray and jet mode feedback. Note that these 50\,$h^{-1}\,{\rm Mpc}$ simulations also share the same initial condition. By comparing these models, we can isolate the impact of individual AGN feedback modes on the SHMR.

Haloes are identified on the fly during the simulation run using a 3D friends-of-friends (FoF) algorithm with a linking length of 0.2 times the mean inter-particle spacing.  The halo masses in this paper are the FoF halo mass, except for Fig.~\ref{fig:1} and Fig.~\ref{fig:6} where we computed $M_{200m}$ in order to be consistent with observational comparisons; this makes little difference in our results. The adoption of FoF halo mass is to avoids pseudo halo growth introduced by the overdensity method\cite{Diemer2013}, which is important for studying evolutionary trends in Section \ref{sec:evolution}. However, there is little difference between the two mass definitions at $z=0$, as shown by the negligible differences between the red and blue galaxy median lines in Fig.~\ref{fig:1} and Fig.~\ref{fig:2}. 

Galaxies are identified using a 6D phase-space galaxy finder within each dark matter halo in the {\tt yt}-based package {\tt Caesar}. Many galaxy properties are computed after the identification.  The galaxy magnitudes in SDSS $g$ and $r$ bands are computed using the {\tt PyLoser} package within {\tt Caesar}, which employs the FSPS stellar population synthesis code\cite{Conroy2009,Conroy2010} to compute spectral energy distributions and includes line-of-sight extinction based on the self-consistently evolved dust content in \simba. We apply a stellar mass cut $M_* \geq 10^{10} M_\odot$ to ensure that we only include well-resolved central galaxies ($\gtrsim 500$ star particles), which in any case covers the same mass range as the observations shown in Fig.~\ref{fig:1}. The simulated galaxies are further separated into red and blue by their $g-r$ colour as in observations\cite{More2011}. We note that the separation line, $(g-r)_{cut} = 0.65 + 0.075 * (\log (M_*/[h^2_{0.71} M_\odot]) - 10)$, has a slightly lower slope compared to \cite{More2011}, because massive galaxies in \simba include the intra-cluster light which has a younger age than the BCG \cite{Cui2014b}. Finally, galaxies and haloes are linked with their progenitors through matching their unique particle IDs.  As we only focus on the central galaxy and its host halo in this study, the merger history is built based on the main progenitors of $z=0$ haloes. If not specified, our results only apply within this resolved mass range ($M_* \gtrsim 10^{10} M_\odot$, or $M_{\rm halo} \gtrsim 10^{11.5} M_\odot$).

\subsection{Halo formation, galaxy transition and jet mode AGN feedback ignition times.} \label{sec:time_definations}

As illustrated in Supplementary Fig. 3, the histories of galaxy and halo formation are rather different, while the BH growth history is more similar to the galaxy growth history.  Individual haloes or galaxies have fairly similar growth histories as the medians.  We use the commonly-adopted half-mass redshift, when the halo's most massive progenitor obtained half of its $z=0$ mass, for the halo formation time.  We have calculated the halo formation time both by interpolation of the data points and from the data smoothed with the Savitzky-Golay filter \cite{Savitzky1964} which has been integrated in {\sc Scipy} \cite{Scipya,Scipyb}.  The two methods give a consistent result.  Galaxy mass accretion histories can be generally characterised by two processes: a fast growing period in its early phase and a constant/quiescent period after the transition.  Therefore, it is natural to choose the connecting point between the two periods as their transition time.  We first fit the galaxy formation history with a step function by joining an error function term (for the fast growing period) and a linear term (for the constant period).  We note here that we simply use the total stellar mass of galaxies as they evolve without considering in-situ/ex-situ growth modes or rejuvenation processes. The advantages of this simple treatment are (1) we do not need a separate explanation for the scatter in the two phases; (2) it is very hard to separate in-situ and ex-situ growth in observations. Then, we get the galaxy transition time $z_{\rm T}$ through the slope of the fitting curve, as the time when
\begin{equation}
    \frac{d \log M_*}{d t [{\rm Gyr}]} < 0.1.
    \label{eq:1}
\end{equation}
We note here that this definition is similar to a threshold in sSFR, but it is based on a much longer time baseline, and it includes stellar mass brought in by mergers.  Our conclusions are not affected by the choice of this threshold, which only produces a systematic shift in the transition time. 

The BH mass growth history can also be roughly separated into a fast growing period in its early phase and a constant period later on. This is consistent with two modes of BH accretion in \simba. Furthermore, this also correlates with the radiative and jet mode AGN feedback, which is dominated in high Eddington ratios in the fast BH mass growing period and in low Eddington ratios ($f_{\rm Edd} < 0.2$) in the constant period, respectively. To directly link the BH mass change with the Eddington ratio, we use equation 11 from \cite{Thomas2019} to correlate the Eddington ratio with the BH specific growth rate, and define the jet mode feedback ignition time as the time when:
\begin{equation}
    \frac{d \log M_\bullet}{d t [{\rm Gyr}]} < 0.9.
    \label{eq:2}
\end{equation}

\section{Data Availability}
The \simba\ simulation snapshots with {\tt Caesar} halo and galaxy catalogues are publicly available at \url{http://simba.roe.ac.uk/}. The processed source data for producing the figures in this paper is available at the author's repository: \url{https://bitbucket.org/WeiguangCui/ms-mhalo-scatter/src/master/}. 

\section{Code Availability}
The \simba\ simulation is run with a private version of {\sc GIZMO}, which is available from the corresponding author upon reasonable request. The galaxy catalogue of the \simba\ simulation is produced by {\tt Caesar}, which is publicly available at \url{https://github.com/dnarayanan/caesar}. The detailed analysis pipeline scripts are available in the author's repository: \url{https://bitbucket.org/WeiguangCui/ms-mhalo-scatter/src/master/}.

%%\end{methods}

%% Put the bibliography here, most people will use BiBTeX in
%% which case the environment below should be replaced with
%% the \bibliography{} command.

% \begin{thebibliography}{10}
% \expandafter\ifx\csname url\endcsname\relax
%   \def\url#1{\texttt{#1}}\fi
% \expandafter\ifx\csname urlprefix\endcsname\relax\def\urlprefix{URL }\fi
% \providecommand{\bibinfo}[2]{#2}
% \providecommand{\eprint}[2][]{\url{#2}}

% \bibitem{Scipyb}
% \bibinfo{author}{Millman, K.~J.} \& \bibinfo{author}{Aivazis, M.}
% \newblock \bibinfo{title}{Python for scientists and engineers}.
% \newblock \emph{\bibinfo{journal}{Computing in Science and Engineering}}
%   \textbf{\bibinfo{volume}{13}}, \bibinfo{pages}{9--12} (\bibinfo{year}{2011}).
% \newblock \urlprefix\url{https://doi.org/10.1109/MCSE.2011.36}.

% \end{thebibliography}

\bibliographystyle{naturemag}
\bibliography{references}

%% Here is the endmatter stuff: Supplementary Info, etc.
%% Use \item's to separate, default label is "Acknowledgements"

\section{Acknowledgements} The authors express their sincere thanks to the anonymous referees for their invaluable comments, suggestions and kind help, without which this work would be incomplete.  We also acknowledge helpful discussions with Marcel van Daalen, Jorryt Matthee, Katarina Kraljic, Daniele Sorini, Nicole Thomas and Ying Zu.  We thank Robert Thompson for developing {\sc Caesar}, and the {\sc yt} team for development and support of {\sc yt}. W.C. acknowledges the support from the China Manned Space Program through its Space Application System. W.C. \& J.A.P. acknowledge support from the European Research Council under grant number 670193 (the COSFORM project). R.D. acknowledges support from the Wolfson Research Merit Award programme of the U.K. Royal Society. W.C. \& R.D. acknowledge support from the STFC AGP Grant ST/V000594/1. W.C. acknowledge support from the China Manned Space Program through its Space Application System. D.A.A. acknowledges support by NSF grant AST-2009687 and by the Flatiron Institute, which is supported by the Simons Foundation. X.Y. acknowledges support from the national science foundation of China (grant Nos. 11833005, 11890692, 11621303). This work used the DiRAC@Durham facility managed by the Institute for Computational Cosmology on behalf of the STFC DiRAC HPC Facility. The equipment was funded by BEIS capital funding via STFC capital grants ST/P002293/1, ST/R002371/1 and ST/S002502/1, Durham University and STFC operations grant ST/R000832/1. DiRAC is part of the National e-Infrastructure.

\section{Author information}
% \author{Weiguang Cui$^{1}$, Romeel Dav\'e$^1$, John A. Peacock$^1$, Daniel Angl\'es-Alc\'azar$^{2,3}$, \& Xiaohu Yang$^4$}
% \begin{affiliations}
% \item Institute for Astronomy, University of Edinburgh, Royal Observatory, Edinburgh EH9 3HJ, UK
% \item Department of Physics, University of Connecticut, 196 Auditorium Road, U-3046, Storrs, CT 06269-3046, USA
% \item Center for Computational Astrophysics, Flatiron Institute, 162 Fifth Avenue, New York, NY 10010, USA
% \item Department of Astronomy, School of Physics and Astronomy, Shanghai Jiao Tong University, Shanghai, 200240, China
%  % \item separate with \verb|\item| commands.
% \end{affiliations}

\subsection{Contributions} W.C. conceived the project. R.D preformed the simulation with contributions from D.A.A. and provided the {\tt Caesar} catalogue. W.C. designed and performed the analysis. W.C., R.D., J.A.P, D.A.A and X.Y. interpreted the results. W.C. wrote the manuscript with contributions from R.D., J.A.P, D.A.A and X.Y.
\subsection{Corresponding author} Correspondence to Weiguang Cui ~(email: weiguang.cui@ed.ac.uk).

\section{Ethics declarations}
\subsection{Competing interests}
The authors declare no competing interests.

\section{Supplement: Colour evolution} \label{sec:s1}
\begin{figure}
\includegraphics[width=\textwidth]{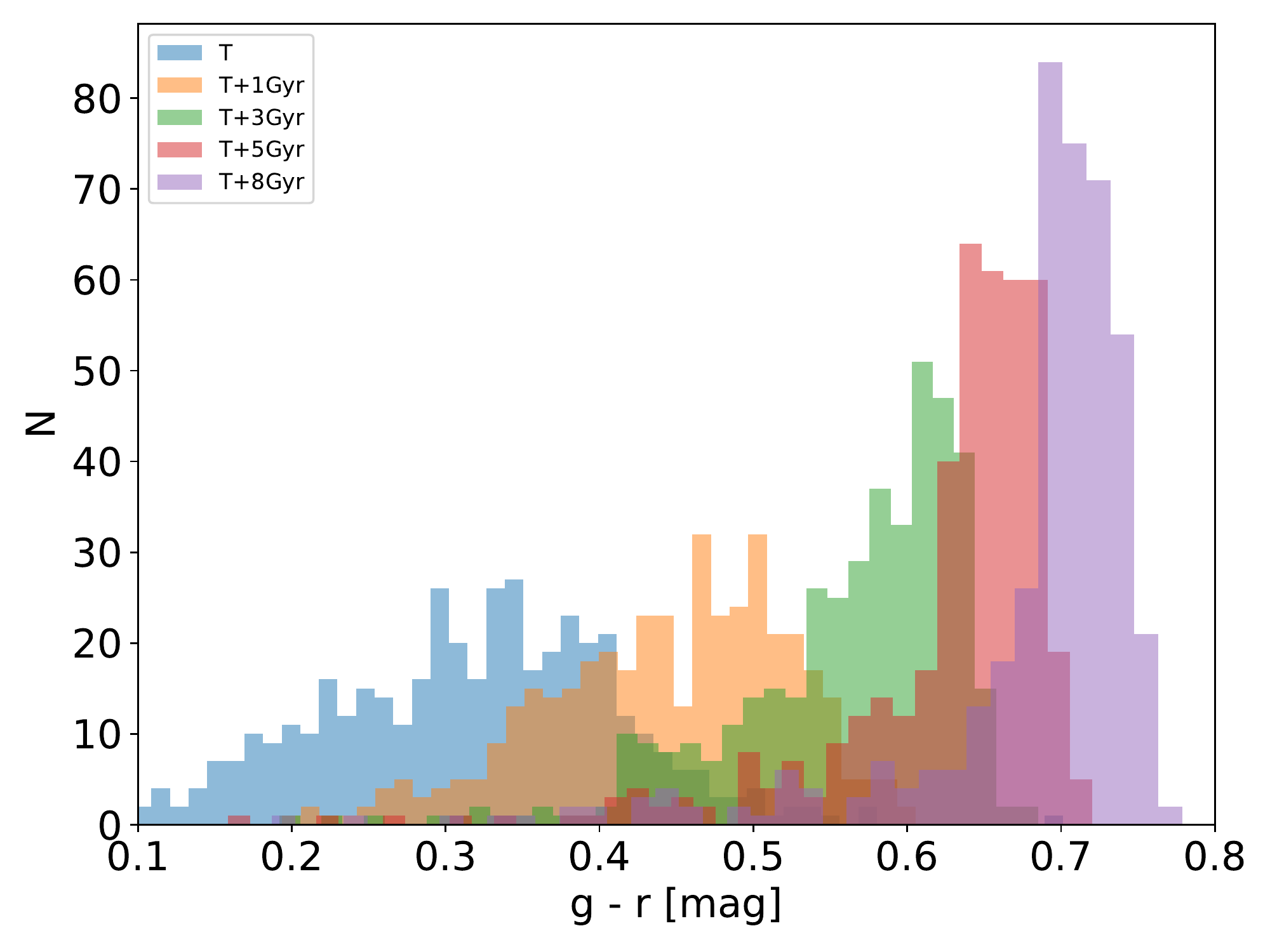} 
\caption{{\bf The $g-r$ colour distributions for galaxies at the transition time and after.} Different colour histograms as indicated in the legend show the galaxy colour distributions at different times. Only galaxies which have their transition time + 10~Gyr less than the Universe age are selected. This plot indicates that the galaxy needs about 8~Gyr after its transition time to reach a red colour ($g - r > 0.65$).}
\label{fig:s1}
\end{figure}

In Fig. 2, we show that the median line from blue galaxies is roughly consistent with the median line for galaxies in transition time bin $0.5 \leq z < 1$ and the magenta line for red galaxies is in agreement with the yellow line from galaxies with their transition time between $z=1-1.5$. It is interesting to know how long the galaxy requires in order to change its colour to red. If the star formation is truncated completely and abruptly, stellar population models suggest this time would be $<2$~Gyr, but this is not what happens in practice. We select the early-transition galaxies that have their transition time when the Universe was less than 3.8 Gyr old. Due to this qualitatively study and a limited sample, we do not further bin the data in stellar mass or redshift. In Supplementary Fig.~\ref{fig:s1}, we show the $g-r$ colour distribution of these galaxies at their transition time and 1, 3, 5, 8~Gyr after the transition time in different colours. To be classified as red galaxies with $g-r > 0.65$, the galaxy needs about 8~Gyr. This estimation is roughly consistent with the results that red galaxies generally have their transition time $z_T>1$; see Fig. 2 for details. We refer to \cite{Chaves-Montero2020} for a detailed study on relation between the galaxy colour with its star formation history.

\section{Supplement: The inversion problem} \label{sec:s2}
\begin{figure}
\includegraphics[width=\textwidth]{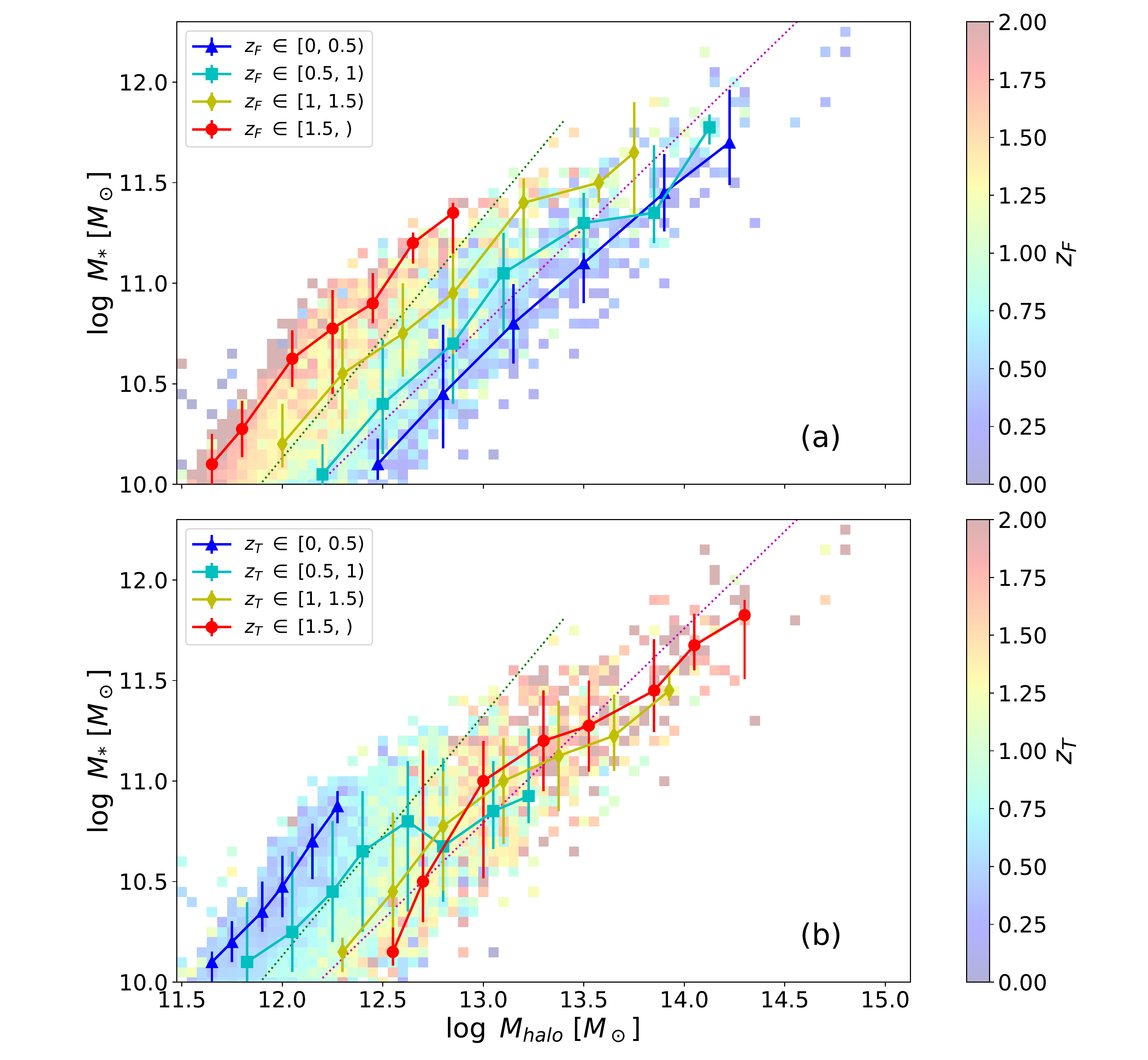} 
\caption{{\bf The SHMR shown in reversed axes.} The same as Fig. 2, but reversed axes with $M_{\rm halo}$ as the $x$-axis.  The median lines are shown binned in halo mass. Note that the same symbols, lines and error bars as in Fig. 2 are adopted. In panel a showing colour-coding by halo formation time, the various $z_{\rm F}$ binned medians do not overlap, showing no inversion problem.  In panel b, the cyan $z_{T}$ median overlaps with the yellow and red median at $M_{\rm halo}>10^{12.8}M_\odot$, showing an inversion problem here.}
\label{fig:s2}
\end{figure}

It has been noted\cite{Moster2018,Moster2020} that binning the data in halo mass vs. in stellar mass can result in reversed trends. It is worth investigating here whether our results are affected by this inversion problem. In Supplementary Fig.~\ref{fig:s2}, we represent the results from Fig. 2 but with swapped axes, so that the binning is done in halo mass instead of stellar mass. For the halo formation time $z_{\rm F}$ in the top panel, it is clear that the median lines for different halo formation times are basically unchanged, i.e. the blue, cyan, yellow, and red lines maintain the same ordering at all halo masses. The lower panel shows that this is also true for the median lines in different galaxy transition time bins for lower halo masses $11.5 < \log M_{\rm halo} \lesssim 12.8$. Therefore, the SHMR from \simba does not have the inversion problem\cite{Moster2018,Moster2020} at least in this halo mass range. Namely, at fixed stellar mass, blue galaxies tend to live in haloes with lower mass while red galaxies tend to have a massive host halo. Conversely, at a given halo mass, galaxies with a higher stellar mass tend to be blue (indicating a higher specific star formation rate), while galaxies having a lower stellar mass tend to be red and quiescent. 
However, in a more massive halo mass range $\log M_{\rm halo} \gtrsim 12.8$, late transition galaxies are comparable to or lie below the early transition galaxies (see the cyan line vs. the red/yellow lines) which indeed shows a reversed trend compared to Fig. 2.  Thus there is potentially an inversion problem here.  However, the sample size is small: there are very few galaxies with late transition times at this halo mass range in \simba. Moreover, our results are driven primarily by galaxy quenching, which occurs within the mass range unaffected by inversion in either $z_{\rm F}$ or $z_{\rm T}$.  Hence our main results are not strongly affected by the inversion problem, at least in the main halo mass range $11.5 < \log M_{\rm halo} \lesssim 12.8$. We would like to further note that our analysis is also limited in galaxy stellar mass. Extending to low stellar mass galaxies, we may experience the inversion problem again as these should be dominated by star forming galaxies.

\section{Supplement: The formation histories} \label{sec:s3}
\begin{figure}
\includegraphics[width=\textwidth]{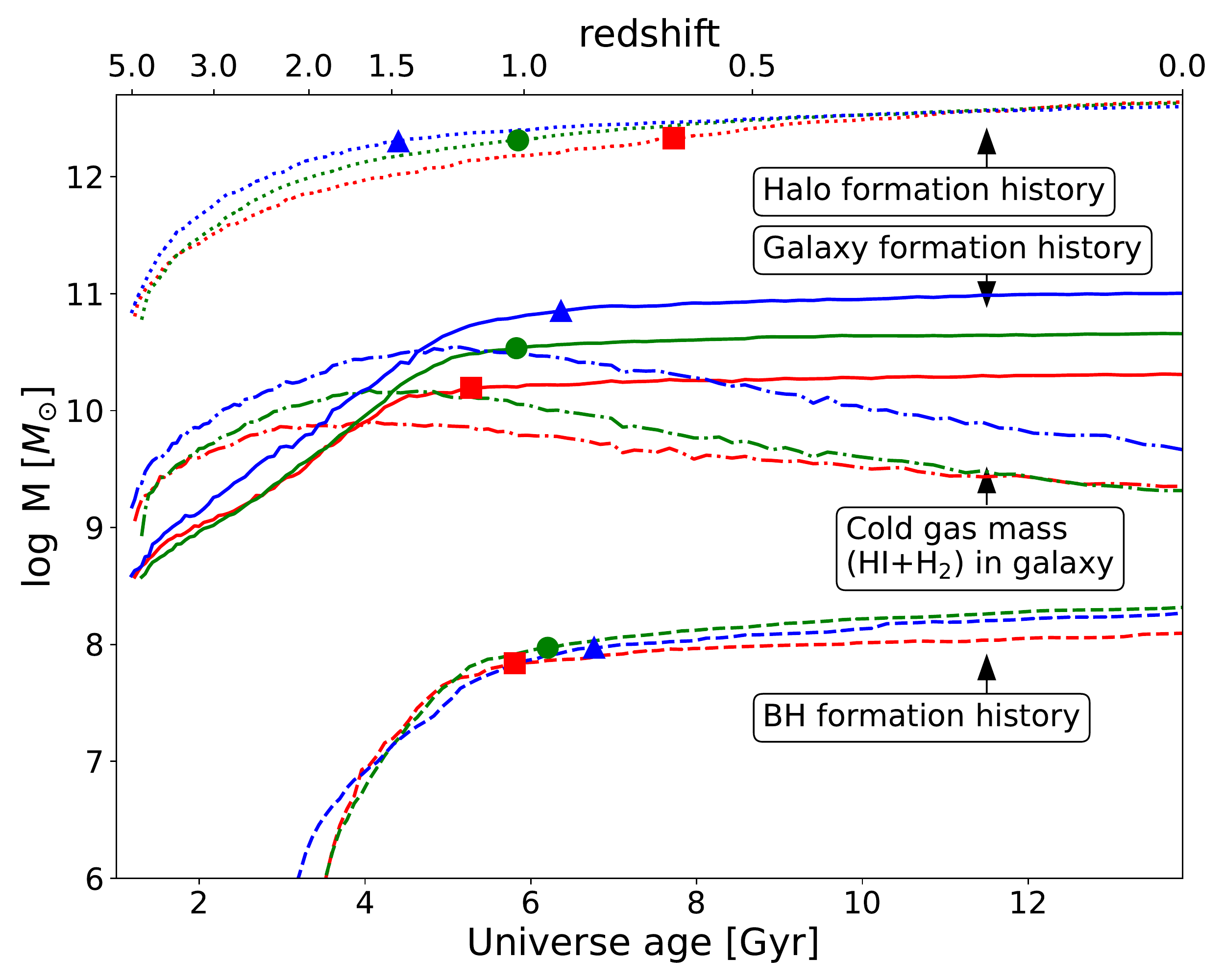} 
\caption{{\bf The mass accretion history of halo, galaxy, cold gas and BH.} The median halo mass (dotted lines), galaxy stellar mass (solid), cold gas mass (dot-dashed) and BH mass (dashed) in galaxies as a function of time/redshift for a sample of haloes with $10^{12.5} < M_{\rm halo} < 10^{12.8}$ at $z=0$. As depicted in Fig. 3, haloes and galaxy are grouped into red, green blue regions, indicated by the correspondingly coloured lines. The stars identify the median halo formation time, galaxy transition time and BH jet mode on time along each of the tracks, respectively.}
\label{fig:s3}
\end{figure}

To set the stage for presenting the evolution of the SHMR and to link to the keys of the galaxy transition, we illustrate the galaxy and halo growth histories in \simba for several key quantities in this study: halo formation time, galaxy transition time, AGN jet mode feedback igniting time and galaxy crossing time (see Methods).  
In Supplementary Fig.~\ref{fig:s3} the ordering of the median times (stars) along the various tracks shows that halo formation time and galaxy transition time are anti-correlated, as we saw earlier.  Additionally, we see that the early forming haloes (blue lines) host more massive galaxies, which have a slightly longer growing period, and hence a later transition time.  We will argue that this is owing to their enhanced cold gas content, defined in \simba as the total neutral gas mass (HI+H$_2$) in the galaxy, at all epochs, which sustains star formation for a longer period. Therefore, to quantify the correlation between gas evolution with the galaxy transition time, we define a galaxy `crossing time' when the stellar mass is equal to its cold gas mass ($t_{M_* = M_{\rm gas}}$); we will explore this time in relation to the transition time in  \S 4.2.  

There are three AGN feedback modes in \simba(see Methods): radiative mode, jet mode and X-ray heating from black holes. Since the Eddington ratio is proportional to the specific growth rate of the BH, the transition from fast BH mass growth (high Eddington ratio) to slow BH mass growth (low Eddington ratio) separates the radiative mode and jet mode AGN feedback. We denote this time as the BH jet-on time ($t_{\rm jet\, on}$).
It is also important to note that the BH jet-on time is quite similar to the galaxy transition time (see \S 4.2). This is not a coincidence: in \simba, BH jets are primarily responsible for eventually quenching galaxies\cite{Dave2019}. Since the jet mode turns on at low Eddington ratios, and accretion becomes slower when there is less cold gas around, this timescale is also correlated with the galaxy crossing time. 
The interplay of these various quantities thus determines the final location of galaxies in the SHMR plane.

\section{Supplement: Discussions on the current state of affairs} \label{sec:s4}

The scatter in SHMR correlates with two key properties -- the galaxy colour and the halo formation time. The former is not easy to model in hydrodynamic simulations, while the later is hard to measure in observation. Therefore, the detailed connections are still not clear and the origin of the scatter in SHMR is still under debate. 

The correlation between the scatter in SHMR and galaxy colour is in reasonably agreement in observation: the higher stellar mass galaxies tend to have a blue colour compared to the lower stellar mass galaxies at the same halo mass~\cite{PostiFall2021}. However, contradiction results on the connection to halo formation time have been reported: Observational work from the Sloan Digital Sky Survey (SDSS) using satellite kinematics\cite{Wojtak2013} showed that red galaxies with a lower stellar mass live in haloes with higher concentrations,  thus purportedly early-formed haloes, compared to blue galaxies at the same halo mass. Conversely, using halo masses from a weak lensing analysis of the SDSS redMaPPer clusters for $M_{\rm halo} \approx 1.74\times 10^{14} M_\odot/h$, it was shown that lower stellar mass galaxies live in lower concentration haloes\cite{Zu2020}. 

On the theoretical side, the connection between the scatter in the SHMR and the halo formation time is found to be in general agreement: the higher stellar mass galaxies tend to live in early-formed haloes compared to the lower stellar mass galaxies at the same halo mass\cite{Matthee2017}. This will lead to an opposite galaxy colour distribution compared to observation, i.e. galaxies with a higher stellar mass would have redder colour, if we naively assume that galaxies in early-formed haloes have more time to accrete and form stars\cite{Zehavi2018}. Similarly, Emerge\cite{Moster2020} also predicted that, at fixed halo mass, passive (red) galaxies have a higher stellar mass than active (blue) galaxies, which is understood as a result of linking SFR and halo growth rate in their model. Similarly, \cite{Montero-Dort2021} found that haloes with a faster accretion of their halo masses at early times, tend to lose their gas faster, reach the peak of their star formation histories at higher redshift, and become quenched earlier.

Using the EAGLE simulation\cite{Schaye2015}, \cite{Matthee2019} suggested that, at fixed stellar mass, galaxies with a higher sSFR (blue galaxies) tend to form later, although this relation is much weaker at $M_* > 10^{10.5} M_\odot$ (Spearman rank correlation coefficient $R_s \approx -0.1$) than at $10^9 < M_* < 10^{10} M_\odot$ ($R_s \approx -0.45$). Through private emails, we also confirmed that, in a fixed halo mass bin, late formed haloes tend to host galaxies with a very weakly higher sSFR (0.05 dex), which tend to be blue and to have a higher stellar mass. However, also using the EAGLE simulation, \cite{Kulier2019} found that the scatter in the SHMR anti-correlates with the mean age of the galaxy stellar population, such that galaxies with a higher stellar mass (which also inhabit early-formed haloes) are younger at a fixed halo maximum circular velocity. Furthermore, \cite{Correa2020} shows that disc galaxies in EAGLE tend to reside in earlier forming haloes than their elliptical counterparts in same-mass haloes. Both seem to contradict to the result from \cite{Matthee2019}, if we assume that galaxies with a younger age and disc have a bluer colour or higher sSFR. It may be that the disagreements come from the different indicators used for blue galaxies (age and halo maximum circular velocity instead of halo mass\cite{Kulier2019} vs. discs\cite{Correa2020} vs. sSFR\cite{Matthee2019}), but this is speculative.

\end{document}